\documentclass[final,1p,times]{elsarticle}
\pdfoutput=1
\usepackage{amsmath}
\usepackage{dsfont}
\usepackage{relsize}
\usepackage{subfig}

\renewcommand{\vec}[1]{\textbf{#1}}
\newcommand{\tensor}[1]{\mathbb{#1}}
\newcommand{\dd}{\mathrm{d}}

\begin{document}

\journal{Journal of Computational Physics}
\begin{frontmatter}
\title{Coupled Vlasov and two-fluid codes on GPUs}
\author{M. Rieke}
\author{T. Trost}
\author{R. Grauer}
\ead{grauer@tp1.rub.de}
\address{Institut f\"ur Theoretische Physik I,
Ruhr-Universit\"at Bochum, 44801 Bochum, Germany}
\begin{abstract}
We present a way to combine Vlasov and two-fluid codes for the
simulation of a collisionless plasma in large domains while keeping
full information of the velocity distribution in localized areas of
interest. This is made possible by solving the full Vlasov equation in
one region while the remaining area is treated by a 5-moment two-fluid
code. In such a treatment, the main challenge of coupling kinetic and
fluid descriptions is the interchange of physically correct boundary
conditions between the different plasma models. In contrast to other
treatments, we do not rely on any specific form of the distribution
function, e.g. a Maxwellian type. Instead, we combine an
extrapolation of the distribution function and a correction of the
moments based on the fluid data.  Thus, throughout the simulation both
codes provide the necessary boundary conditions for each other. A
speed-up factor of around 20 is achieved by using GPUs for the
computationally expensive solution of the Vlasov equation and an
overall factor of at least 60 using the coupling strategy combined
with the GPU computation. The coupled codes were then tested on
the GEM reconnection challenge.
\end{abstract}

\begin{keyword}
coupling, Vlasov equation, multifluid, GPU, reconnection, GEM
\end{keyword}

\end{frontmatter}

\section{Introduction}

For many relevant problems in plasma physics, the plasma can be
considered non-collisional or even collisionless. In these cases,
the Vlasov-Maxwell system is sufficient for a complete description of
the physically important phenomena.  Nevertheless, it is close to
impossible to numerically solve the Vlasov equation for global
problems on fluid scales due to the vast amount of memory needed to
resolve a 6-dimensional phase-space. However, in many situations a
complete kinetic description is not necessary and computationally
inexpensive fluid models based on the moments of the distribution
function are sufficient.  In addition, plasmas are typical multi-scale
phenomena. Often there are only localized and small
regions, where a kinetic description is needed, whereas the rest of
the domain can be described by means of a fluid model. This
separation of scales can be found in many typical problems of plasma
physics, such as reconnection in the magnetotail of the earth, solar
flares or typical phenomena in tokamaks. To determine what amount of
computational savings coupling techniques might provide, a rough
estimation of the numerical costs can be made on the basis of spatial
scales. To do this, both the computational effort for the fluid region
and the overhead for the coupling itself can be assumed as negligible
compared to the costs of the kinetic part of the simulation. Thus the
fraction of the domain where kinetic effects occur can indicate
how much computational time might be saved. This fraction
can be estimated on the basis of typical plasma parameters. The most
relevant scales are determined by the ion skin depth, below which the
dynamics of ions and electrons decouple, and the ion gyro radius, which
is especially important in the context of non-collisional plasmas.

In the magnetotail of the earth, kinetic effects are important in and
around the current sheet, where reconnection takes place. In an example
of a typical setup for this problem given by \citet{bir2001}, the
current sheet--and thus the kinetic region--covers around 10\% of the
computational domain.  In a typical tokamak with an overall size $\sim
1\,\mathrm{m}$, the ion skin depth is $\sim 5\,\mathrm{cm}$ (see
\citet{woods2006}). The ion gyro radius is even one order of magnitude
smaller. Thus, taking into account the rotationally symmetric geometry
of a tokamak and the magnetic field therein, the fraction of the
domain where localized kinetic effects as reconnection might take
place is likely to scale with the fraction of $5\,\mathrm{cm}/
1\,\mathrm{m}$ and can thus be expected to be only 5\% of the domain,
to give a very rough estimation of the order of magnitude. In a
solar flare, the ratio is even smaller: A typical size of $\sim
10^7\,\mathrm{m}$ compares to an ion skin depth of about
$10\,\mathrm{m}$ (see for example \citet{ram2005}).

If the kinetic regions are known, this additional information might be
used to reduce the computational costs drastically by restricting the
expensive kinetic simulation to this smaller area and using a suitable fluid
model for the rest of the domain. There have been several approaches
to exploit this ansatz in numerical schemes. In the works of
\citet{deg2010}, \citet{del2003}, \citet{gou2013}, \citet{kla2000},
and \citet{Tal1997} methods for coupling the Boltzmann equation to
some kind of fluid model like the Navier-Stokes or Euler-equations
are adressed. \citet{Sug2007} describe the coupling of PIC and MHD
models, \citet{Mar2014} show a way to combine PIC and fluid models,
and \citet{Kol2012} describe the transition from neutral gas models to
models of weakly ionized plasmas with respect to coupling kinetic and
fluid equations. \citet{Schulze2003} use coupling techniques in the
context of epitaxial growth.

Most of these schemes are based on the
strong assumption that the distribution function is close to
Maxwellian in the vicinity of the coupling border and use this approximation 
to generate boundary conditions for the kinetic region. In other
approaches, an artificial field is introduced that expresses the
deviation from a Maxwellian and is coupled to the underlying equations
\cite{deg2010}, or the fluxes at the boundary are used for the coupling
\cite{Tal1997}.

This paper presents a different mechanism for coupling the
Vlasov equation with a five-moment model and the full Maxwell
equations. Our method is easy to implement and does not rely
on any specific assumptions with respect to the distribution function
near the coupling border. In addition, no extensive transition region
between the kinetic and fluid domains is required.
 
The rest of this paper is organized as follows: After the physical models have
been described in Sec.~\ref{sec:models}, the numerical schemes 
are presented in Sec.~\ref{sec:numerics}.  The coupling
procedure is explained in Sec.~\ref{sec:coupling}. Finally, in
Sections~\ref{sec:results}~and~\ref{sec:summary}, first numerical
results are presented and discussed.

\section{Physical models}
\label{sec:models}
Consider a plasma consisting of different particle species $s$. The
Vlasov equation
\begin{equation}
\label{eq:vlasov}
\partial_tf_s + \vec{v}\cdot\nabla_{\vec x}f_s + 
\frac{q_s}{m_s}(\vec E + 
\vec v\times \vec B)\cdot\nabla_{\vec v}f_s = 0
\end{equation}
describes the time evolution of the phase-space density $f_s(\vec x,
\vec v, t)$, where $\vec E$ is the electric field, $\vec B$ is the
magnetic field, $q_s$ is the charge, and $m_s$ is the mass per
particle. By calculating the moments of $f_s$ with respect to $\vec
v$, we can derive important physical quantities describing the plasma:
\begin{flalign}
\label{eq:mass-density}
&\text{the mass density} & \rho_s &= m_s\int f_s \dd^3v, &\\
\label{eq:momentum-density}
&\text{the momentum density} & \vec u_s &= m_s\int \vec v f_s \dd^3v, &\\
\label{eq:energy-density}
&\text{and the energy density} & \mathcal E_s &= \frac{m_s}{2}\int \vec v^2 f_s \dd^3v. &
\end{flalign}
Their time dependence can be calculated from the Vlasov equation
\eqref{eq:vlasov} and is given by:
\begin{align}
\partial_t\rho_s &= -\nabla\cdot\vec u_s \\
\partial_t\vec u_s &= -\nabla\cdot\left(\frac{\vec u_s\otimes\vec u_s}{\rho_s} + 
\tensor{P}_s\right) + \frac{q_s}{m_s}(\rho_s\vec E + \vec u_s\times\vec B_s)\\
\partial_t\mathcal{E}_s &= -\nabla\cdot\left(\frac{\mathcal{E}_s\mathds{1} + 
\tensor{P}_s}{\rho_s}\vec u_s\right) -
\nabla\cdot\vec{Q}_s + \frac{q_s}{m_s}\vec u_s\cdot\vec E,
\end{align}
where $\tensor{P}_s$ is the pressure tensor, $3p_s =
\mathrm{Tr}\;\tensor{P}_s$ is the trace of the pressure tensor, and
$\vec{Q}_s$ is the heat flux tensor.  These quantities must be
provided in order to close the system.

For our first tests of the
coupling procedure, we assume that the plasma is adiabatic and the
pressure to be isotropic, which corresponds to
\begin{equation}
\tensor{P}_s = p_s\mathds{1},\hspace{1em}p_s = \frac{2}{3}\mathcal{E}_s -
\frac{1}{3}\frac{\vec{u}_s^2}{\rho_s},\hspace{1em}\text{and}\hspace{1em}\nabla\cdot\vec{Q}_s = 0
\end{equation}
respectively. Note that this fluid closure was chosen for simplicity
and other fluid models may be used depending on the situation of the plasma
application.

Furthermore Faraday's and Amp\`ere's laws
\begin{align}
\label{eq:faraday}
\partial_t\vec B &= -\nabla\times\vec E \\
\label{eq:ampere}
\partial_t\vec E &= c^2(\nabla\times\vec B - \mu_0\vec j)
\end{align}
are required to determine how the electromagnetic fields evolve where
current density $\vec j$ is given by
\begin{equation}
\vec j = \sum_s\frac{q_s}{m_s}\vec u_s.
\end{equation}

\section{Numerical schemes}
\label{sec:numerics}

In order to solve the coupling problem, we must first design
numerical schemes for solving the Vlasov equation, the five-moment
two-fluid equations and the Maxwell equations separately. It is important to note
that we have chosen the fluid model with care such that we do not
impose a generalized Ohm's law to obtain the electric field $\vec{E}$
but use the same set of Maxwell equations as for the Vlasov equations.
This implies that timescales of both fluid and kinetic descriptions
are comparable. This allows for a clean separation between the Maxwell
solver, the Vlasov solver and the fluid solver. That is, the Maxwell
solver only requires the current density as a source term, which can be
obtained from the fluid and the Vlasov solver.

Coupling to even larger spatial and temporal scales which could be
described with an MHD model, would thus be more related to the
question of coupling Maxwell's equation to a generalized Ohm's law.
This kind of coupling can be done on the fluid level (2-fluid with
Maxwell-equation coupled to MHD with generalized Ohm's law). This is
still a major task and will be the next necessary step.

Both models, the Vlasov equation and the two-fluid equations, are
implemented numerically and each code can be executed alone or
coupled to the other---a concept known as \emph{multiple program
  multiple data} (MPMD).
  
Because of the high dimensionality of the distribution function $f_s$,
the Vlasov code is much more computationally expensive than the fluid
code. Therefore, to achive optimal performance, the Vlasov code is
executed on multiple GPUs, while the relatively low cost fluid model
and Maxwell's equations \eqref{eq:faraday} and \eqref{eq:ampere} are
solved on CPUs. The network communication is implemented via OpenMPI.
Our code solves the 2.5-dimensional problem (i.e. two space dimensions
and three velocity dimensions), but a generalisation to the full
3-dimensional case is straightforward. In the following subsections we
will describe the numerical schemes. In addition to that, a detailed
description of semi-Lagrangian solvers is given because they are
essential both to the Vlasov code and to the coupling routine that
will be explained in section \ref{sec:coupling}.

\subsection{Numerical implementation of the Vlasov equation}

In this section, the numerical solver for the Vlasov equation used in
this work is described. By means of
splitting methods, the Vlasov equation is broken down into a set of
simpler one-dimensional conservation laws. These can then be
solved with a semi-Lagrangian scheme. Such a scheme requires finding
the characteristics for which we use the Backsubstitution method, and
approximating the integrals between these characteristics, which we do
by means of the PFC scheme.

\subsubsection{Conservative semi-Lagrangian schemes}
The underlying idea of how to solve conservation laws
by means of semi-Lagrangian schemes is now described. As an example consider
\begin{equation}
\label{eq:conservation-law}
\partial_tf(x,t) + \partial_x\big(u(x,t)f(x,t)\big) = 0.
\end{equation}
Let the domain be divided into equal cells and the $i^{\mathrm{th}}$
cell be bounded by the points $x_{i}$ and $x_{i+1}$. Different points
in time, separated by the time interval $\Delta t$, are denoted as $t^n
= n\Delta t$. The scheme will then evolve the cell integrals
\begin{equation}
\overline{f_i}(t^n) = \int_{\mathlarger{x_i}}^{\mathlarger{x_{i+1}}}f(x,t^n)\,\dd x
\end{equation}
in time. Next, a family of curves $\mathcal{C}(t;x,s)$ is introduced, called the
\emph{characteristics} of equation \eqref{eq:conservation-law}, which
are defined by
\begin{align}
\frac{\dd}{\dd t}\mathcal{C}(t) &= u(\mathcal{C}(t),t)\\
\mathcal{C}(t=s) &= x.
\end{align}
Integrating $f$ along $x$ between the two curves that pass through
the cell boundaries of cell $i$ at time $t^{n+1}$, taking the time
derivative, using the chain rule for parameter integrals, and
inserting the conservation law \eqref{eq:conservation-law} leads to 
\begin{equation}
\frac{\dd}{\dd t}
\int_{\mathlarger{\mathcal{C}(t;x_i,t^{n+1})}}^{\mathlarger{\mathcal{C}(t;x_{i+1},t^{n+1})}}f(x,t)\,\dd x = 0.
\end{equation}
If the velocity field $u(x,t)$ is Lipschitz continuous, the
Picard-Lindel\"of theorem will guarantee that two different
characteristics will never intersect. Then we may integrate over time
to obtain
\begin{equation}
\label{eq:update}
\overline{f_i}(t^{n+1}) = 
\int_{\mathlarger{\mathcal{C}(t^n;x_i,t^{n+1})}}^{\mathlarger{\mathcal{C}(t^n;x_{i+1},t^{n+1})}}f(x,t^n)\,\dd x.
\end{equation}
Where $\overline{f_i}(t^n)$ and the points
$\mathcal{C}(t^n;x_i,t^{n+1})$ are known, the cell integrals
$\overline{f_i}(t^{n+1})$ can be found by reconstructing the right
hand side of equation \eqref{eq:update} from the $\overline{f_i}(t^n)$
by means of an arbitrary reconstruction scheme.  In other words, in
order to solve the conservation law \eqref{eq:conservation-law}, it is
sufficient to find the source points of the characteristics that will
pass through the cell boundaries.

\subsubsection{Splitting}
The phase space density $f_S$ depends on six coordinates plus
time. Trying to solve the Vlasov equation in a naive way by simply
discretizing it based on an arbitrary scheme would lead to severe
problems: If the number of dimensions is $d$, the number of direct
neighbours of a given cell is $3^d-1$. Even if one spatial dimension
can be left out because of the symmetry of the problem, this still
results in 242 cells. A stencil spanning a larger part of these cells,
would produce a very clumsy and slow scheme. This exponential growth
of the number of dependencies can be avoided by splitting the Vlasov
equation into smaller parts that are more manageable and can be solved
separately. If this is done carefully, the resulting error does not
exceed the one of the overall scheme.

The Vlasov equation can be
written in the form
\begin{equation}\label{eq:vlasov-operator}
  \partial_t f_s + (\mathcal{A} + \mathcal{B})f_s = 0
\end{equation}
with operators
\[ \mathcal{A} = \vec{v}\cdot\nabla_{\vec x}\]
and
\[ \mathcal{B} = \frac{q_s}{m_s}(\vec E + 
\vec v\times \vec B)\cdot\nabla_{\vec v}. \]
Instead of solving the full equation \eqref{eq:vlasov-operator}, we consider the equations
\begin{equation}\label{eq:vlasov-a}
  \partial_t f_s + \mathcal{A}f_s = \partial_t f_s + \vec{v}\cdot\nabla_{\vec x}f_s = 0
\end{equation}
and
\begin{equation}\label{eq:vlasov-b}
  \partial_t f_s + \mathcal{B}f_s = \partial_t f_s + \frac{q_s}{m_s}(\vec E + 
\vec v\times \vec B)\cdot\nabla_{\vec v}f_s = 0.
\end{equation}
Now, given some differential equation
\begin{equation}\label{eq:general-operator}
  \partial_t f + \mathcal{S}f = 0, \quad f( t = 0 ) = f_0
\end{equation}
we can write its formal solution as $f(\Delta t) = \exp(-\Delta t\mathcal{S})f_0 =:
T(\mathcal{S})f_0$.

With this notation in mind we can decompose the solution of
\eqref{eq:vlasov-operator} in terms of solutions of
\eqref{eq:vlasov-a} and \eqref{eq:vlasov-b} by making use of the
Strang splitting method:
\begin{equation}
  f_s( t + \Delta t ) = T(\mathcal{A}+\mathcal{B})f(t) = T\left(\frac{1}{2}\mathcal{A}\right)T(\mathcal{B})T\left(\frac{1}{2}\mathcal{A}\right)f(t) + \mathcal{O}(\Delta t^3).
\end{equation}
This decomposition is exact if the operators $\mathcal{A}$ and
$\mathcal{B}$ commute. However, this is not the case here, but
the resulting error is of third order in time and thus does not
destroy the overall accuracy.
Now that the problem has been split into two subproblems, each one can be
solved separately. In both \eqref{eq:vlasov-a} and \eqref{eq:vlasov-b}
the operators still act on a three-dimensional space. It is natural to
try to decompose the operators $\mathcal A$ and $\mathcal B$
further in order to simplify the problem even more and make it
possible to solve it with a semi-Lagrangian scheme which only
works for one dimension. As for $\mathcal A$, which is the
sum of three commuting operators, this splitting is exact and can
easily be done by a simple composition of the single operators.

The
operator $\mathcal{B}$, however, does not allow for such an easy
decomposition. Here, it would be possible to use Strang splitting
again, but there would be some disadvantages in this. First, the
resulting scheme would be non-isotropic, as the single operators would
not be used the same number of times. Second, the application of these
operators is particularly expensive, so it is not desirable to apply
them no more than necessary.

Thus, the next problem is to find a splitting so that
equation \eqref{eq:vlasov-b} can be solved by only three applications
of the time-evolution operator (one for each dimension of $\vec v$)
and obtain at least second-order accuracy in space and time. We
rewrite \eqref{eq:vlasov-b} as
\begin{equation}
  \partial_t f_s + \mathcal{B}f_s = \partial_t f_s + (\mathcal{V}_x + \mathcal{V}_y + \mathcal{V}_z)f_s = 0
\end{equation}
with
\[ \mathcal{V}_i = \frac{q_s}{m_s}( E_i + (\vec{v}\times\vec{B})_i)\partial_i\]
and make the ansatz
\begin{equation}\label{eq:ansatz-splitting-v}
f(t+\Delta t) = T(\tilde{\mathcal{V}}_z)T(\tilde{\mathcal{V}}_y)T(\tilde{\mathcal{V}}_x)f(t) + E_\mathrm{split} + E_\mathrm{mod}.
\end{equation}
with modified operators $\tilde{\mathcal{V}}_i$ that still have to be
determined. The term $E_\mathrm{split}$ represents the error due to
the splitting scheme, while $E_\mathrm{mod}$ is the error caused by
modifying the operators. If the modifications are chosen correctly,
both error terms cancel out and the splitting is exact. It should be 
noted however, that the error term of the splitting depends on the
order in which the solution operators are applied. Thus it is not
possible to change the order of execution of the
$T(\tilde{\mathcal{V}}_i)$ once the modifications have been chosen.
The implementation of \eqref{eq:ansatz-splitting-v} is done by means of
the Backsubstitution method, which is described in the next
section.

\subsubsection{The Backsubstitution Method}

The Backsubstitution method \cite{schmitz1,schmitz2} relies on the
calculation of the characteristics needed for the semi-Lagrangian
schemes that occur in the single operators in
\eqref{eq:ansatz-splitting-v} by means of the Boris push method. The
latter is best known from the context of PIC simulations.

We start with a review of the Boris push method. Considering the
discretized equation
\begin{equation}
\label{eq:implicit-lorentz}
\frac{\vec{v}^{n+1} - \vec{v}^n}{\Delta t} = \frac{q_\alpha}{m_\alpha}
\left(\vec{E}^{n+1/2} + \frac{\vec{v}^{n+1} + \vec{v}^n}{2}\times \vec{B}^{n+1/2}\right),
\end{equation}
the effects of the electric and magnetic fields can be separated by introducing
new variables
\begin{equation}
\label{eq:new-v-variables}
\vec{v}^+ = \vec{v}^{n+1} - \frac{q_\alpha}{m_\alpha}\frac{\Delta t}{2}\vec{E}^{n+1/2}
\hspace{1cm}\mathrm{and}\hspace{1cm}
\vec{v}^- = \vec{v}^{n} + \frac{q_\alpha}{m_\alpha}\frac{\Delta t}{2}\vec{E}^{n+1/2}.
\end{equation}
With that, equation \eqref{eq:implicit-lorentz} becomes
\begin{equation}
\frac{\vec{v}^{+} - \vec{v}^-}{\Delta t} = \frac{q_\alpha}{m_\alpha}
\frac{\vec{v}^{+} + \vec{v}^-}{2}\times \vec{B}^{n+1/2}.
\end{equation}
This equation is solved in two steps. Introducing the quantities
\begin{equation}
\vec{t} = \frac{q_\alpha}{m_\alpha} \frac{\Delta t}{2} \vec{B}
\hspace{1cm}\mathrm{and}\hspace{1cm}
\vec{s} = \frac{2\vec{t}}{1 + t^2},
\end{equation}
the first step calculates an intermediate vector $\vec{v}'$ from $\vec{v}^-$, the 
second step uses this intermediate vector to calculate the result $\vec{v}^+$
\begin{align}
\label{eq:bsf1}
\vec{v}' &= \vec{v}^- + \vec{v}^- \times \vec{t}\\
\label{eq:bsf2}
\vec{v}^+ &= \vec{v}^- + \vec{v}' \times \vec{s}
\end{align}
There is also a backward-in-time version of the Boris push that calculates $\vec{v}^-$ from
$\vec{v}^+$:
\begin{align}
\label{eq:bsb1}
\widetilde{\vec{v}}' &= \vec{v}^+ - \vec{v}^+ \times \vec{t}\\
\label{eq:bsb2}
\vec{v}^- &= \vec{v}^+ - \widetilde{\vec{v}}' \times \vec{s}.
\end{align}
A big advantage of the Boris push is that it correctly conserves
energy respectively the length of $\vec v$ if only magnetic and no
electric fields are present.

Following the idea of cascade interpolation \cite{leslie-purser}, the
first step of the update in velocity space will be performed along
the $\vec{v}_x$ direction, the second step along $\vec{v}_y$, and the
last step along $\vec{v}_z$. For the last step, the distance along the
$\vec{v}_z$ direction is just the $z$ component of the characteristic
ending exactly at the grid point.  Since we are tracing the
characteristics back in time, the grid-point coordinates are denoted
by $\vec{v}^{n+1}$. By using the new variables
\eqref{eq:new-v-variables}, the electric fields can be removed from
the equations and the grid-point coordinates become $\vec{v}^+$.  The
distance for the $\vec{v}_z$ update is then given by a formula $v^-_z
= v^-_z(v^+_x, v^+_y, v^+_z)$.  This notation means that $v_z^-$ is a
function of the variables $v^+_x$, $v^+_y$, and $v^+_z$.  Using the
$z$ component of equations \eqref{eq:bsb1} and \eqref{eq:bsb2}, one
obtains
\begin{equation}
\label{eq:bsz}
v_z^- = v_x^+(t_zs_x - s_y) + v_y^+(t_zs_y + s_x) + v_z^+(1-t_xs_x-t_ys_y)
\end{equation}
for the sought-after function. The step along the $\vec{v}_y$
direction follows a characteristic passing through $v_z^+$ at time
$t-\Delta t$  and $v_y^+$ and $v_x^+$ at time $t$. This means that an
equation of the form $v_y^- = v_y^-(v_x^+, v_y^+, v_z^-)$ is needed.
This can be obtained by solving Eq.~\eqref{eq:bsz} for $v_z^+$ and
inserting the result into the $y$ component of equation
\eqref{eq:bsb2}, yielding
\begin{align}
\label{eq:bsy}
v_y^- &= v_x^+\left(s_xt_y+s_z - \frac{(t_zs_x-s_y)(t_ys_z-s_x)}{1-t_xs_x-t_ys_y}\right)\\
\nonumber &\;\;\;+ v_y^+\left(1-t_xs_x-t_zs_z-\frac{(t_zs_y + s_x)(t_ys_z-s_x)}{1-t_xs_x-t_ys_y}\right)\\
\nonumber &\;\;\;+ v_z^-\frac{t_ys_z-s_x}{1-t_xs_x-t_ys_y}.
\end{align}
For the last step, the equation for $v_z^-$ has the form $v_z^- = v_z^-(v_x^+, v_y^-, v_z^-)$. In the
original version of the backsubstitution method \cite{schmitz1}, this equation was obtained in a 
similar way by solving Eqs. \eqref{eq:bsz} and \eqref{eq:bsy} for $v_z^+$ and $v_y^+$, respectively, 
and ``substituting them back'' into the $x$ component of equation \eqref{eq:bsb2}. However, this
results in a lengthy calculation, and it is difficult to simplify the resulting expression. There
is an easier way to obtain the final result using the forward-in-time version of the Boris push
and solving the $x$ component of equation \eqref{eq:bsf2} for $v_x^-$, which results in
\begin{equation}
\label{eq:bsx}
v_x^- = \frac{v_x^+ - v_y^-(t_xs_y + s_z) - v_z^-(t_xs_z - s_y)}{1-t_ys_y - t_zs_z}.
\end{equation}
Equations \eqref{eq:bsx}, \eqref{eq:bsy}, and \eqref{eq:bsz} provide the starting points for the one-dimensional
updates along the $\vec{v}_x$, $\vec{v}_y$, and $\vec{v}_z$-axis, respectively. The order in which these
steps must be applied is $\vec{v}_x$, then $\vec{v}_y$, then $\vec{v}_z$.

\subsubsection{The PFC Scheme}\label{sssec:pfc}

In this section we present a brief review of the positive flux-conservative (PFC)
scheme developed by \citet{filbet2001conservative}, which is
used in this work and provides a method for calculating
the integral in Eq.~\eqref{eq:update}.  For a comparison of other
schemes that might be used for the same purpose, see
\citet{crouseilles2010conservative}.
  
The PFC scheme is used to approximate the function $f$ on the basis of
its cell integral $f_i$ in the $i^\mathrm{th}$ cell. Here, the case of
a positive velocity $u$ is discussed. The case $u<0$ can be derived
analogously.
From Eq.~\eqref{eq:update} it holds that
 \begin{equation}
\label{eq:recall-cell-integral}
 {f_i^n} = \int_{x_{i-1/2}}^{x_{i+1/2}}f(x,t^n)\dd x =
 \int_{\mathlarger{X(t^{n-1};x_{i-1/2},t^{n})}
 }^{\mathlarger{X(t^{n-1};x_{i+1/2},t^{n})}}
 f(x,t^{n-1})\dd x.
 \end{equation}
 By setting
 \begin{equation}
   \label{eq:flux-definition}
   \Phi_{i+1/2}=\int_{\mathlarger{X(t^{n-1};x_{i+1/2},t^{n})}
   }^{\mathlarger{x_{i+1/2}}}f(x,t^{n-1})\dd x,
 \end{equation}
equation \eqref{eq:recall-cell-integral} can be rewritten as an update equation for the $f_i$:
 \begin{equation}
   \label{eq:pfc-update}
 f_i^n = f_i^{n-1} + \Phi_{i-1/2} - \Phi_{i+1/2}.
 \end{equation}
 The primitive function of $f$ at the cell borders is given by
 \begin{equation}
 F(t^{n}, x_{k-1/2}) = \sum_{i=0}^{k-1}f_i^n.
 \end{equation}
 An approximation $\widetilde F$ of $F$ within the $i^\mathrm{th}$ cell
 can be obtained by Newton interpolation of $F$ on the cell borders
 $\{x_{i-3/2}, x_{i-1/2}, x_{i+1/2}, x_{i+3/2}\}$:
 \begin{align}
 \widetilde F(x) =\; &F(x_{i-3/2}) + f_{i-1}(x-x_{i-3/2}) +
 \frac{f_i-f_{i-1}}{2\Delta x} (x-x_{i-3/2})(x-x_{i-1/2})\\ \nonumber &+
 \frac{f_{i+1} - 2 f_i + f_{i-1}}{6(\Delta x)^2}
 (x-x_{i-3/2})(x-x_{i-1/2})(x-x_{i+1/2})
 \end{align}
 \normalsize
 By differentiating and rearranging the terms, one obtains an
 approximation $\widetilde f$ in the $i^\mathrm{th}$ cell: \small
 \begin{align}
 \widetilde f(x) &= \frac{\dd \widetilde F(x)}{\dd x}(x)\\
 &= f_i + \frac{f_{i+1}-f_i}{6(\Delta x)^2}\big(2(x-x_i)(x-x_{i-3/2}) +
 (x-x_{i-1/2})(x-x_{i+1/2})\big)\\ \nonumber
 &\hspace{1em}+\frac{f_i-f_{i-1}}{6(\Delta x)^2}\big(2(x-x_i)(x-x_{i+3/2})
 + (x-x_{i-1/2})(x-x_{i+1/2})\big)
 \end{align}
 \normalsize 
This reconstruction can create negative values of $\widetilde f(x)$, or
values larger than $f_\infty = \max(f(x))$, which would be an unphysical
solution to the Vlasov equation. Thus, slope limiters are introduced:
 \begin{align}
 \epsilon_i^+ &= \begin{cases}
 \min(1; 2f_i/(f_{i+1}-f_i)) & \mathrm{if\;} f_{i+1} > f_i\\
 \min(1; -2(f_\infty - f_i)/(f_{i+1}-f_i)) & \mathrm{if\;} f_{i+1} < f_i
 \end{cases}\\
 \epsilon_i^- &= \begin{cases}
 \min(1; 2(f_\infty - f_i)/(f_i-f_{i-1})) & \mathrm{if\;} f_i > f_{i-1}\\
 \min(1; -2f_i/(f_i-f_{i-1})) & \mathrm{if\;} f_i < f_{i-1},
 \end{cases}
 \end{align}
 and the approximation becomes
 \small
 \begin{align}
 \widehat f_i(x) = f_i &+ \epsilon_i^+\frac{f_{i+1}-f_i}{6(\Delta x)^2}\big(2(x-x_i)(x-x_{i-3/2}) +
 (x-x_{i-1/2})(x-x_{i+1/2})\big)\\ \nonumber
 &+\epsilon_i^-\frac{f_i-f_{i-1}}{6(\Delta x)^2}\big(2(x-x_i)(x-x_{i+3/2})
 + (x-x_{i-1/2})(x-x_{i+1/2})\big).
 \end{align}
 \normalsize The flux $\Phi_{i+1/2}$ can be approximated by
 integrating $\widehat f_i(x)$ from $X(t^{n-1};x_{i+1/2},t^{n})$ to
 $x_{i+1/2}$ (see equation \eqref{eq:flux-definition}). It was assumed
 that the velocity $u$ used in calculating the characteristic $X$ is
 positive. Because the characteristic is followed backwards in time,
 its footpoint lies to the left of $x_{i+1/2}$.  However, due to the
 semi-Lagrangian nature of this scheme, the point
 $X(t^{n-1};x_{i+1/2},t^{n})$ might not be located in the
 $i^\mathrm{th}$ cell, but multiple cells further to the left. Setting
 $j$ so that $X(t^{n-1};x_{i+1/2},t^{n})$ lies in the $j^\mathrm{th}$
 cell and defining $\alpha_i = (x_{j+1/2} -
 X(t^{n-1};x_{i+1/2},t^{n}))/\Delta x$ one obtains
 \begin{align}
 \Phi_{i+1/2} = \sum_{k=j+1}^{i}f_k + \alpha_i 
 \left(\vphantom{\frac{\epsilon_j^+}{6}}f_j\right.
 &+ \frac{\epsilon_j^+}{6}(1-\alpha_i)(2-\alpha_i)(f_{j+1}-f_j)\\ \nonumber
 &+ \left.\frac{\epsilon_j^-}{6}(1-\alpha_i)(1+\alpha_i)(f_j-f_{j-1})\right).
 \end{align}
 For negative velocities, $\alpha_i = (x_{j-1/2} -
 X(t^{n-1};x_{i+1/2},t^{n}))/\Delta x$ can be used in
 \begin{align}
 \Phi_{i+1/2} = -\sum_{k=j+1}^{i}f_k + \alpha_i 
 \left(\vphantom{\frac{\epsilon_j^+}{6}}f_j\right.
 &- \frac{\epsilon_j^+}{6}(1-\alpha_i)(1+\alpha_i)(f_{j+1}-f_j)\\ \nonumber
 &- \left.\frac{\epsilon_j^-}{6}(2+\alpha_i)(1+\alpha_i)(f_j-f_{j-1})\right).
 \end{align}
 If the $\alpha_i$ are known for each cell, these formulas together with
 equation \eqref{eq:pfc-update} provide the PFC scheme. 

\subsection{Numerical implementation of the two-fluid model}

For the discretization in space, we use the CWENO scheme
\cite{kur2000}, an easy to implement third order finite-volume scheme.
The third order strong-stability-preserving Runge-Kutta scheme by
\citet{shu88} is employed for the time integration.

\subsection{Maxwell solver}

The electromagnetic fields live on a Yee grid \cite{yee1966}. They are
evolved through the FDTD method presented in \cite{taf1975}. Since the
speed of light exceeds all other speeds found in the plasma by far,
subcycling is used in order to resolve lightwaves while keeping the
global timestep as large as possible.

To obtain the electromagnetic fields and the sources at the correct
positions, we use a linear interpolation. This results in a second
order accurate scheme which is consistent with the Vlasov code.

\section{Coupling}
\label{sec:coupling}


The three separate codes, the Vlasov code, the multifluid code, and
the Maxwell code have to be coupled in order to complete our
multiphysics approach to plasma simulations.
The Maxwell solver is mostly independent of the other two schemes and
can be solved globally in the complete domain. As a
source term, it requires only the local current density, which can easily be calculated
from the phase space distribution functions or its first moments.
Similarly, the electromagnetic fields appear as
(space and time dependent) source terms in the fluid and
Vlasov equation and are directly provided by the Maxwell solver. Thus,
no restrictions with respect to the domain or the other two schemes
are implied by the way in which Maxwell's equations are solved. There
is a further advantage to this approach: Because of the
far-reaching separation between the Maxwell solver and the other
schemes, the global timestep is not limited by the speed of light.
During each step of the fluid solvers, several subcycling steps of the
Maxwell scheme can be carried out in order to resolve light waves. If
the interplay is organized correctly, the resulting overall
error is still of second order in the global time step. As each step of the
Vlasov scheme is very expensive, this separation of time scales is
quite desirable and efficient.

The numerical fluid and Vlasov solvers presented above do not
require any non-local operations in physical space. This makes it
possible to subdivide the physical domain into arbitrary pieces or
blocks, in which the equations can be solved independently. Over one
time step, each block has to only provide the boundary conditions for
its neighbouring blocks. In principle, this makes it possible to mix
fluid and kinetic blocks in an arbitrary way as long as meaningful
boundary conditions can be provided. Providing boundary data for the
fluid solver from kinetic data is simple and can be achieved
by calculating the necessary moments via integration of the phase
space density. The difficulty comes in provisioning the boundary
conditions for the Vlasov solver at the boundary to the fluid region.
Even if an arbitrarily high (yet finite) number of moments is known,
there is be no unique way of deriving a correct phase space
density. The necessary information is simply missing and can only be
guessed on the basis of further assumptions. Our approach here is to
extrapolate the shape of the distribution function at the border of
the kinetic region into the cells, where the new distribution is
needed. This extrapolated phase space density is then slightly altered
such that it has the correct lowest moments, which are known from the
fluid region. Thus, the only assumption we enforce is that the
shape of the distribution function does not change too rapidly near
the coupling border. There are neither any assumptions with respect to
any specific form of the distribution function, nor is there any need
for an extensive transition region. In addition, this coupling method
is not limited to the simple fluid model shown above and a huge class
of fluid models may be used in this spirit.

In addition, because electrons and ions are treated separately and are
only coupled via the currents and electromagnetic fields, no
quasi-neutrality is required in the overall scheme.

\subsection{Boundary data for the two-fluid regions}

In the Vlasov regions, the full phase-space densities $f_s$ are known
and the boundary data needed by the two-fluid code can be calculated
by means of the moment equations
\eqref{eq:mass-density}--\eqref{eq:energy-density}.  The Runge-Kutta
scheme that is used for updating the fluid data from time $t$ to time
$t+\Delta t$ needs boundary conditions at times $t$, $t+\Delta t/2$
and $t+\Delta t$. Thus the Vlasov data has to be advanced before the
fluid data. Then, by linear interpolation in time, the intermediate
boundary conditions can be calculated and the fluid simulation can be
advanced.

\subsection{Boundary data for the Vlasov regions}

To update the Vlasov data from $t$ to $t+\Delta t$, boundary
conditions are required only at the initial time $t$. The
full phase-space density is required though and the moments provided by the
fluid code are in general not sufficient to determine $f_s$. If,
however, the coupling boundary is placed in a region that changes
slowly with respect to typical phenomena of the plasma, then it is
appropriate to assume that the shape of $f_s$ will not change
significantly over the distance of a few grid cells---otherwise the
simulation would be underresolved. We may then construct the
phase-space density in the boundary cells by first extrapolating $f_s$
from the Vlasov data and then modifying its shape such that it matches
the moments given by the fluid data. As the extrapolation of the
distribution function to the boundary cells is straightforward, we now
focus on a detailed description of the procedure to adjust the
moments of the distribution function.

\begin{figure}
  \begin{center}
    \includegraphics[width=.9\textwidth]{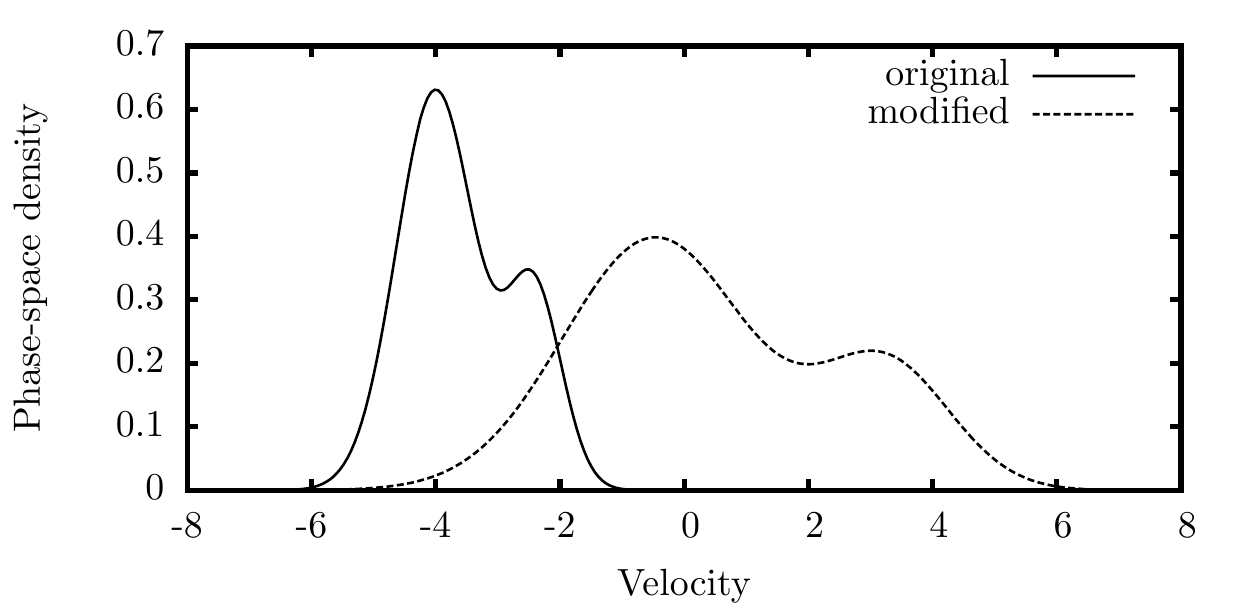}
  \end{center}
  \caption{
  Example of the modification of a distribution function. The solid
  line represents the original distribution function obtained by
  extrapolation. After multiplication with an appropriate factor and
  advection by a ``velocity field'' $a v+ b$, the area under the curve,
  the expectation value and the variance have changed, but the overall
  shape of the distribution function is preserved. Note that this plot
  is exaggerated for clarity.
  }
  \label{fig:fitting}
\end{figure}

To simplify notation, we will drop the subscript $s$ for the particle
species in this section and denote the extrapolated distribution
functions by $f^E$ and the modified functions -- the ones that will be
used as boundary conditions -- by $f^M$. The modification of the
zeroth moment can be performed by a simple \textit{multiplication}.
The first moment can be changed by \textit{shifting} the distribution
function in velocity space. Finally, the second moment is altered by
\textit{squeezing} or \textit{stretching} of the function.
This fitting can be done in various ways. However, here we use the
same semi-Lagrangian method as in the Vlasov solver itself with an
appropriately chosen advecting ``velocity'' field for changing the
distribution function. Note that the method is not limited to a
special choice of a semi-Lagrangian scheme for solving the advection
problem. An exaggerated example is depicted in Fig.~\ref{fig:fitting}.
This procedure will now be explained in
more detail. For simplicity, only the one-dimensional case will be
discussed here. The extension to three dimensions can be obtained by
successive application of the algorithm along each direction. It is
convenient to use slightly different moments than the ones defined in
\eqref{eq:mass-density}--\eqref{eq:energy-density}. Writing
\begin{equation}
n:=\int f(v)\,\dd v,\hspace{2em}V:=\frac{1}{n}\int vf(v)\,\dd v,\hspace{2em}\text{and}\hspace{2em}
T:=\frac{1}{n}\int v^2f(v)\,\dd v - V^2,
\end{equation}
no generality is lost and both sets of moments
can be converted into one another:
\begin{alignat}{3}
\rho(n,V,T) &= mn, &\hspace{1cm} u(n,V,T) &= mnV, &\hspace{1cm} 
\mathcal{E}(n,V,T) &= \frac{m}{2}n\big(T + V^2\big), \\
\label{eq:conversion}
n(\rho,u,\mathcal{E}) &= \frac{\rho}{m}, 
& V(\rho,u,\mathcal{E}) &= \frac{u}{\rho}, & T(\rho,u,\mathcal{E})
&= \frac{2\mathcal{E}}{\rho} - \frac{u^2}{\rho^2}.
\end{alignat}
Now, the idea is to let the distribution function be advected by a 
divergent velocity field $av + b$, i.e. to solve the initial value
problem
\begin{align}
\label{eq:advection}
\partial_\tau f(v,\tau) + \partial_v\Big(\big(av+b\big)f(v,\tau)\Big)
 = 0,\quad f(\tau=0) = f^E,
\end{align}
where $\tau$ is just a dummy time variable independent of the actual
time of the simulation $t$. Note that a rescaling of $\tau$ can
be compensated by a rescaling of $a$ and $b$. Since this equation has the form of a
conservation law, the zeroth moment $n$ of $f$ will not be changed.
Furthermore it will preserve the main features of $f$, like the
number of maxima etc. To solve equation \eqref{eq:advection}
using a semi-Lagrangian scheme, we need to find the source points
of the characteristics.

First we investigate the effects of this advection on the first
moment:
\begin{align}
&\:\partial_\tau\frac{1}{n}\int vf\:\mathrm{d}v + \frac{1}{n}\int v\partial_v\left((av+b)f\right)\:\mathrm{d}v \\
\label{eq:dtV}=&\: \partial_\tau V - b - aV = 0
\end{align}
Here, partial integration and the vanishing of $f$ at infinity is used.
The resulting differential equation can be solved by variation of
constants to yield
\begin{equation}
V(\tau) = \begin{cases} \left(V(0) + \frac{b}{a}\right)\exp(a\tau) - \frac{b}{a} & \text{for}\quad a \in \mathbb{R}\setminus \{0\} \\
V(0) + b\tau & \text{for}\quad a = 0.
\end{cases}
\end{equation}
Solving for $b$ yields
\begin{equation}
\label{eq:b}
b = \begin{cases} \mathlarger{\frac{a\left(V(\tau) - V(0)\exp(a\tau)\right)}{\exp(a\tau)-1}} & \text{for}\quad a \in \mathbb{R}\setminus \{0\} \\[1em]
V(\tau) - V(0) & \text{for}\quad a = 0.
\end{cases}
\end{equation}
Next, we examine the evolution of the second moment $T$:
\begin{align}
&  \quad \partial_\tau\frac{1}{n}\int v^2f\:\mathrm{d}v + \frac{1}{n}\int v^2\partial_v\big((av+b)f\big)\:\mathrm{d}v \\
=& \quad \partial_\tau T + \partial_\tau V^2 - 2bV - 2aT - 2a V^2 \\
=& \quad \partial_\tau T - 2aT = 0.
\end{align}
The solution of the resulting differential equation is
\begin{equation}
T(\tau) = T(0)\exp(2a\tau),
\end{equation}
which can be solved for $a$, giving
\begin{equation}
\label{eq:a}
a = \frac{1}{2}\ln{\left(\frac{T(\tau)}{T(0)}\right)}.
\end{equation}
We now want to find the characteristics $\mathcal{C}$ of equation \eqref{eq:advection}, which can be calculated from
\begin{equation}
\frac{\dd}{\dd \tau} \mathcal{C}(\tau) = a\mathcal{C}(\tau) + b,
\end{equation}
with the initial condition $\mathcal{C}(\tau=0) = \mathcal{C}(0)$. The solution is
\begin{equation}
\mathcal{C}(\tau) = \begin{cases} \left(\mathcal{C}(0) + \frac{b}{a}\right)\exp\big(a(\tau-1)\big) - \frac{b}{a} & \text{for}\quad a \in \mathbb{R}\setminus \{0\} \\
b(\tau-1)+\mathcal{C}(0) & \text{for}\quad a = 0 \; .
\end{cases} 
\end{equation}
Inserting $a$ \eqref{eq:a} and $b$ \eqref{eq:b} yields:
\begin{equation}
\label{eq:sourcepoints}
\boxed{
\mathcal{C}(\tau) = V(\tau) + \sqrt{\frac{T(\tau)}{T(0)}}\big(\mathcal{C}(0) - V(0)\big).
}
\end{equation}
As expected, all explicit dependence on $\tau$ has disappeared.\\
Finally, a summary of the algorithm for changing $f^E$ with moments $\rho^E$, $u^E$, and $\mathcal{E}^E$
to $f^M$ with the prescribed moments $\rho^M$, $u^M$, and $\mathcal{E}^M$:
\begin{enumerate}
\item Calculate $\widetilde{f} = f^E\rho^M/\rho^E$ to change the zeroth moment.\\
  Now $\widetilde{f}$ has moments $\widetilde{\rho}=\rho^M$, $\widetilde{u} = u^E\rho^M/\rho^E$,
  and $\widetilde{\mathcal{E}} = \mathcal{E}^E\rho^M/\rho^E$.
\item Use the formulas \eqref{eq:conversion} to obtain $\widetilde{V}$, $\widetilde{T}$, $V^M$, and $T^M$.
\item Set $V(0) = \widetilde{V}$, $T(0) = \widetilde{T}$, $V(\tau) = V^M$, and $T(\tau) = T^M$. Use a 
  conservative Semi-Lagrangian scheme with the source points of the characteristics given by equation
  \eqref{eq:sourcepoints} to obtain $f^M$.
\end{enumerate}

\subsection{Time stepping}

\begin{figure}
  \begin{center}
    \includegraphics[width=.7\textwidth]{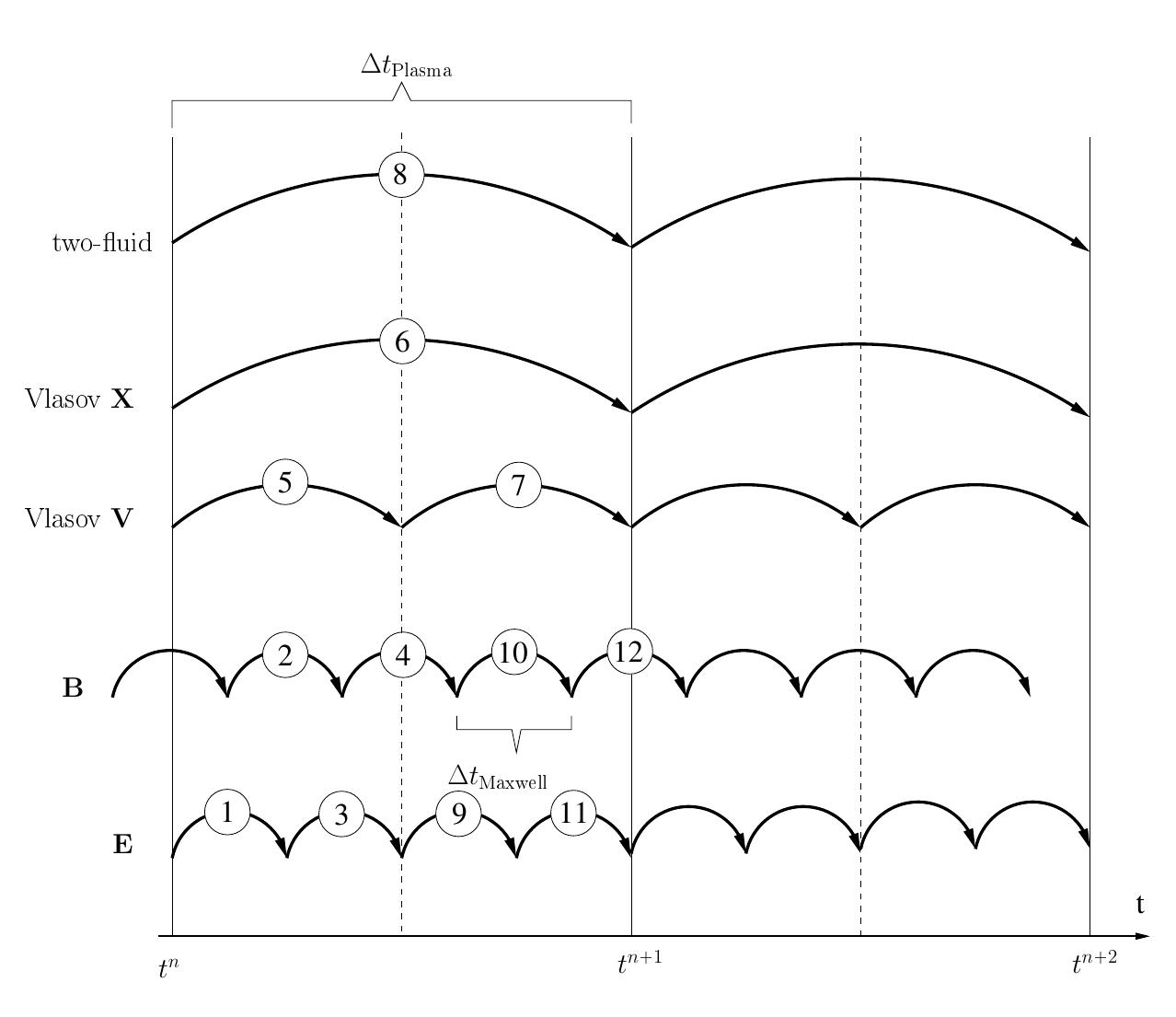}
  \end{center}
  \caption{ This is the order, in which the fields are updated in
    each step. Each arrow represents a substep of the respective
    scheme. The first line (two-fluid) represents the stepping of the fluid quantities. The update of the phase space density in physical (Vlasov $\vec X$)  and
    velocity space (Vlasov $\vec V$) is separated by means of Strang
    splitting. The electric ($\vec E$) and magnetic ($\vec B$) field are updated separately.
  }
  \label{fig:timestepping}
\end{figure}

As multistage and splitting methods are employed, care must be taken
with respect to the points in time at which the fields and the fluxes
at the cell faces live. Fig.~\ref{fig:timestepping} reveals the
structure of time stepping for the electric $\vec{E}$ and magnetic
$\vec{B}$ fields, the fluid quantities and the distribution functions.
The timestep for the Maxwell solver is chosen to be smaller than the global
timestep in order to resolve light waves which are much faster than
all other speeds in the plasma. We choose the ratio between the
smaller timestep of the Maxwell solver and the global timestep as a
power of two, so the electromagnetic fields are known at all relevant
times and the two different discrete times stay aligned throughout the
simulation.

Initially, all the fields exist at the same point in
time. Based on the currents calculated at that time, half a substep of
the magnetic field is done in order to start the scheme. Afterwards,
the following steps are repeated in a cyclic way:

\begin{enumerate}
\item Calculate currents from moments and phase space density of the
  different species and pass them to the Maxwell solver.
\item With these currents as source terms, advance the electromagnetic
  fields by half a global timestep. The magnetic field is shifted by
  half a time step, so linear interpolation is used for obtaining the
  magnetic field at the times at which the electric fields live.
\item Do a complete step of the Vlasov solver, which translates into
  half a step in velocity space, a full step in physical space and
  half a step in velocity space again. Herein, the new electromagnetic
  fields are used as source terms. The boundary conditions for the
  kinetic region are known from the last step.
\item Do a full step of the multi-fluid scheme. Once
  again, the newly obtained electromagnetic fields are used as source
  terms. The Runge-Kutta method requires boundary conditions at
  intermediate times, which must be obtained by temporal interpolation
  of the values provided by the Vlasov solver. Because of this, it is
  necessary to advance the plasma in the kinetic region first.
\item Now the currents can be calculated again. 
\item With these, the
  electromagnetic field is advanced by another half step.
\end{enumerate}

\section{First Results of the multiphysics approach}
\label{sec:results}
As a first test, the GEM reconnection challenge setup, set forth in
\cite{bir2001}, was used. It was simulated both with the multifluid
and the Vlasov code alone, as well as with our multiphysics coupling method.

\begin{figure}
  \begin{center}
    \includegraphics[width=.8\textwidth]{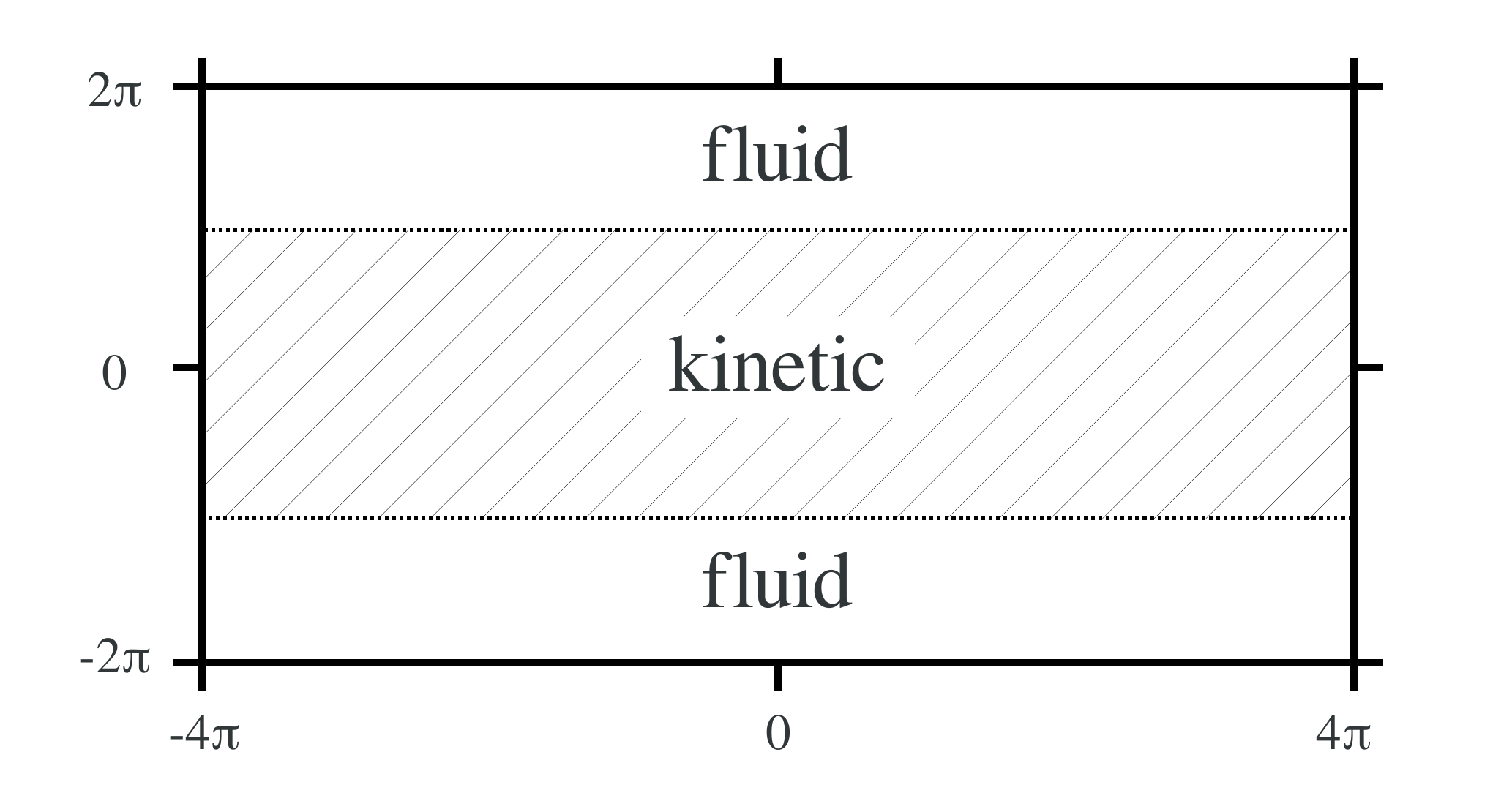}
  \end{center}
  \caption{
  Partition of the domain of the GEM problem. The inner region
  (hatched), i.e. the vicinity of the current layer, is simulated by
  means of the kinetic model, whereas the multifluid code is employed
  in the outer regions.
  }
  \label{fig:partition}
\end{figure}

For respecting the aforementioned restrictions, the vicinity of the
current sheet was completely simulated by means of the Vlasov model
and only in the outer regions the multifluid model was applied (see
Fig.~\ref{fig:partition}). Nevertheless, this already leads to nearly
a half of the required computation time, as only half of the domain
needs to be simulated with a kinetic code and the time needed for the
multifluid code is negligible compared to that for the Vlasov code.\\
A comparison of the time needed for the respective simulations can be
found in Tab.~\ref{tab:speed-up}.\par

\begin{figure}
  \begin{center}
    \includegraphics[width=.8\textwidth]{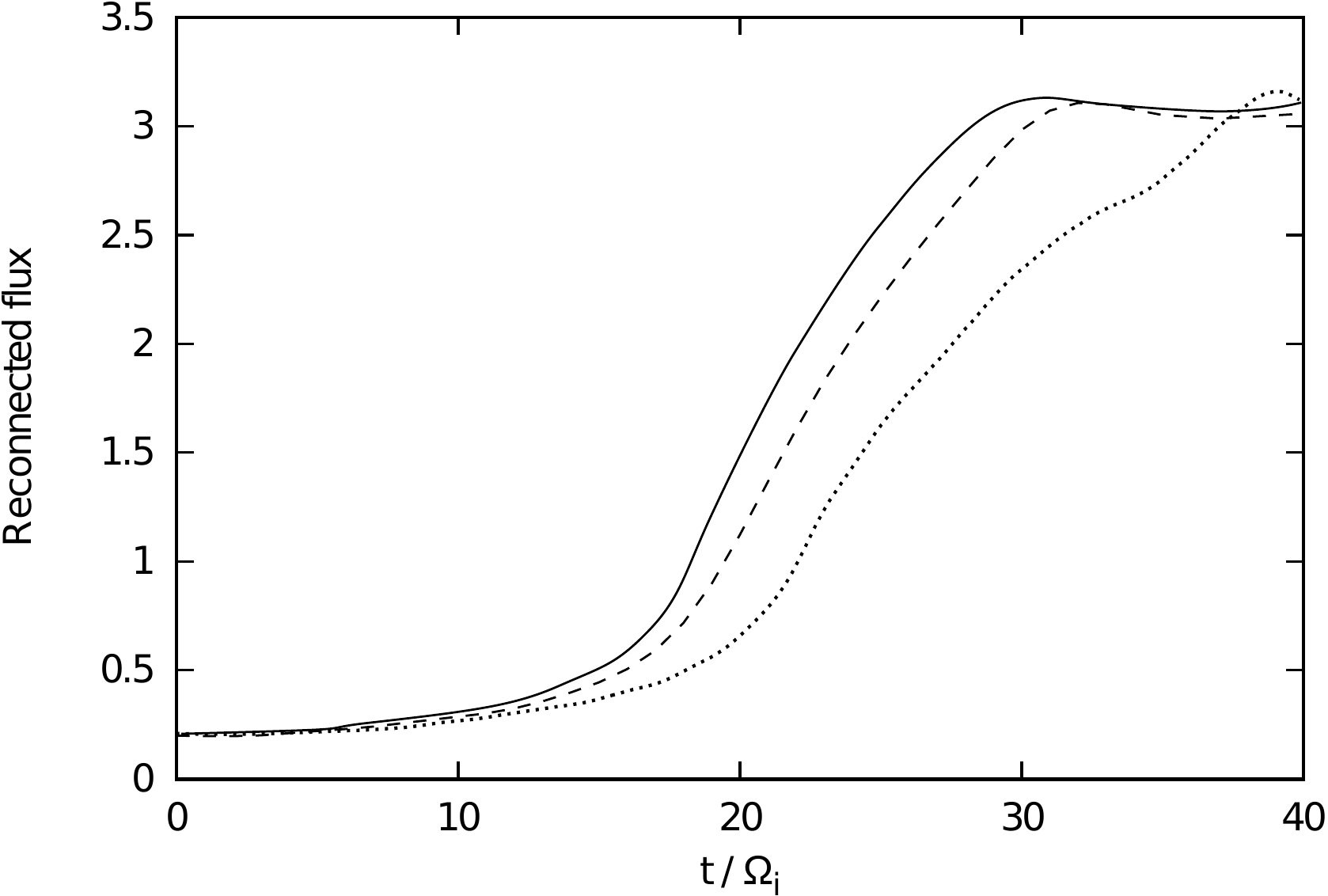}
  \end{center}
  \caption{Comparison of reconnected flux vs.{} time for full kinetic (solid, taken from \cite{sch2006}), coupled (dashed), and pure multifluid (dotted) simulations. }
  \label{fig:gem-curve}
\end{figure}

\begin{figure}
  \begin{center}
    \subfloat[][$j_z$ at $t=20\Omega_i^{-1}$ (pure Vlasov)]{\includegraphics[trim=100mm 30mm 40mm 20mm, clip, width=.49\textwidth]{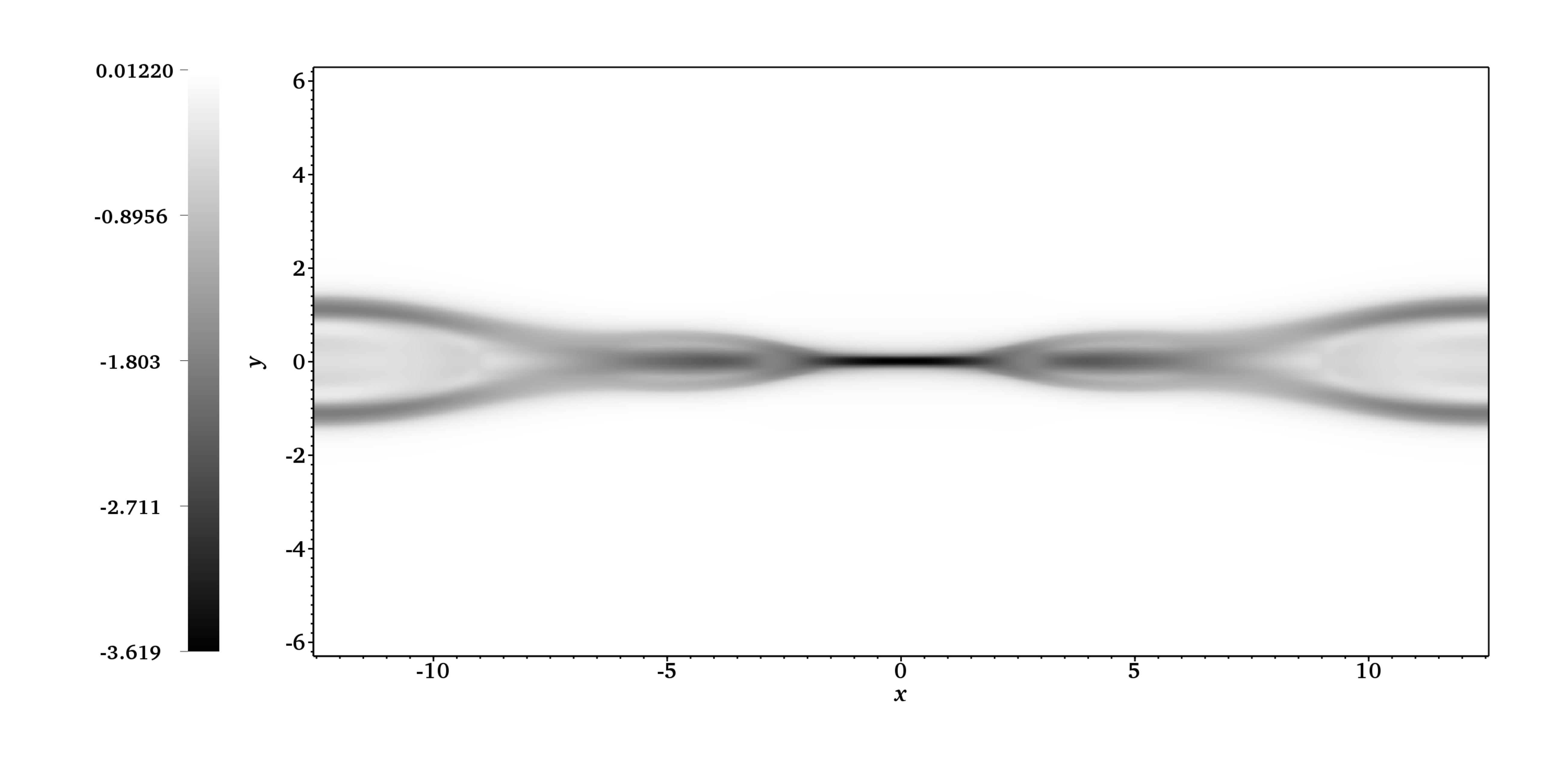}} \subfloat[][Magnetic fieldlines at $t=20\Omega_i^{-1}$ (pure Vlasov)]{\includegraphics[trim=100mm 30mm 40mm 20mm, clip, width=.49\textwidth]{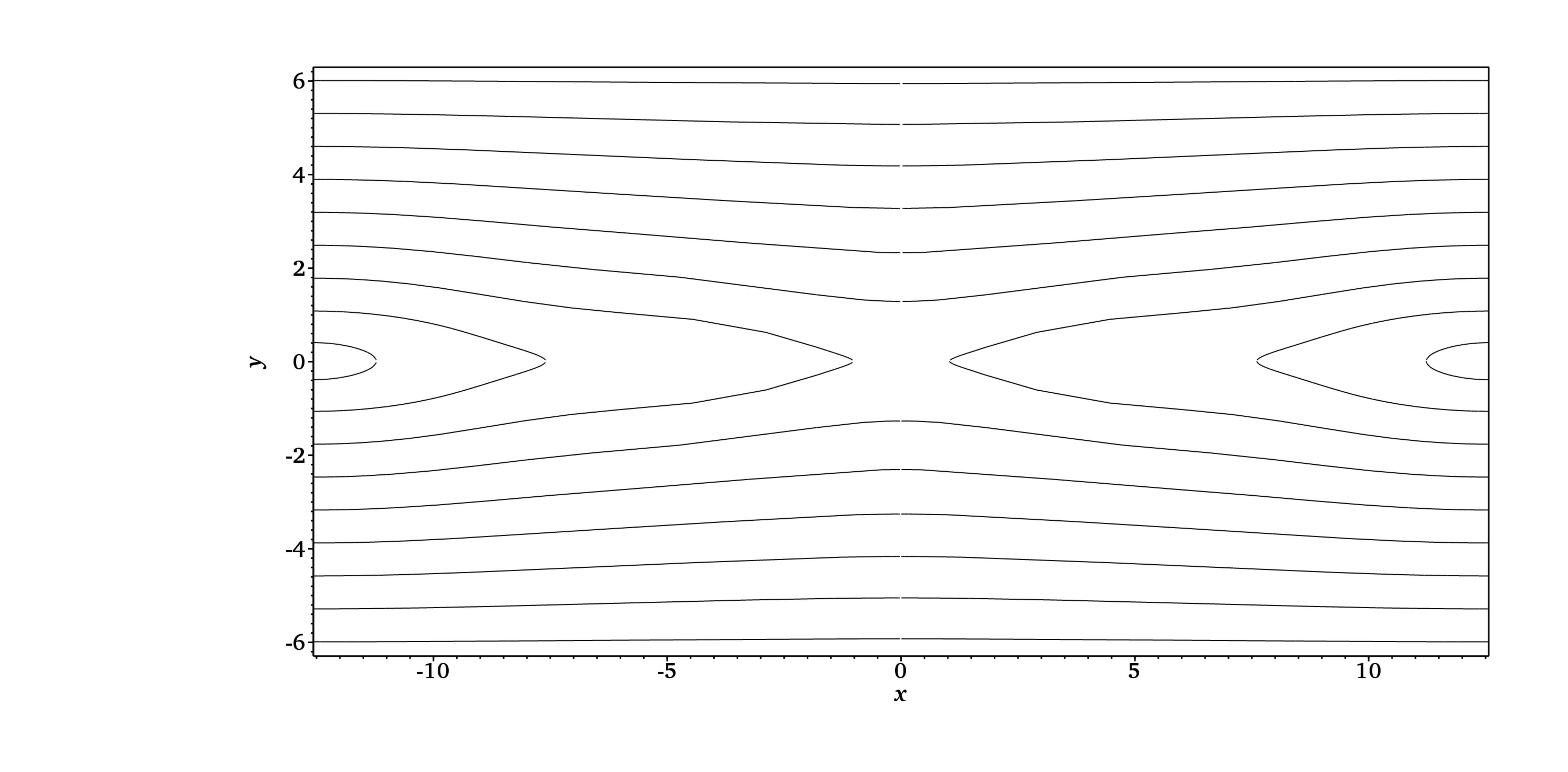}}\\
    \subfloat[][$j_z$ at $t=20\Omega_i^{-1}$ (coupled)]{\includegraphics[trim=100mm 30mm 40mm 20mm, clip, width=.49\textwidth]{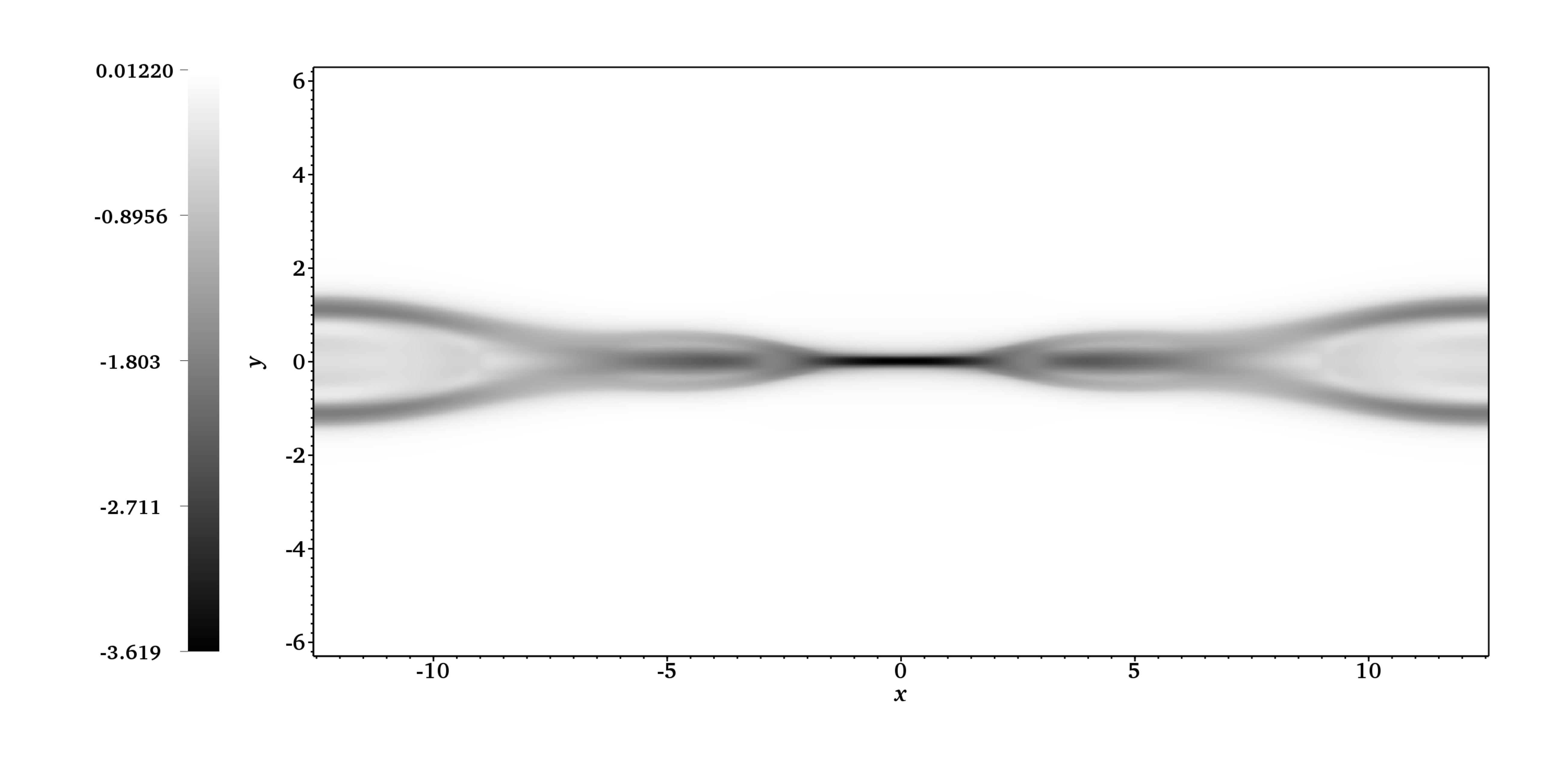}} \subfloat[][Magnetic fieldlines at $t=20\Omega_i^{-1}$ (coupled)]{\includegraphics[trim=100mm 30mm 40mm 20mm, clip, width=.49\textwidth]{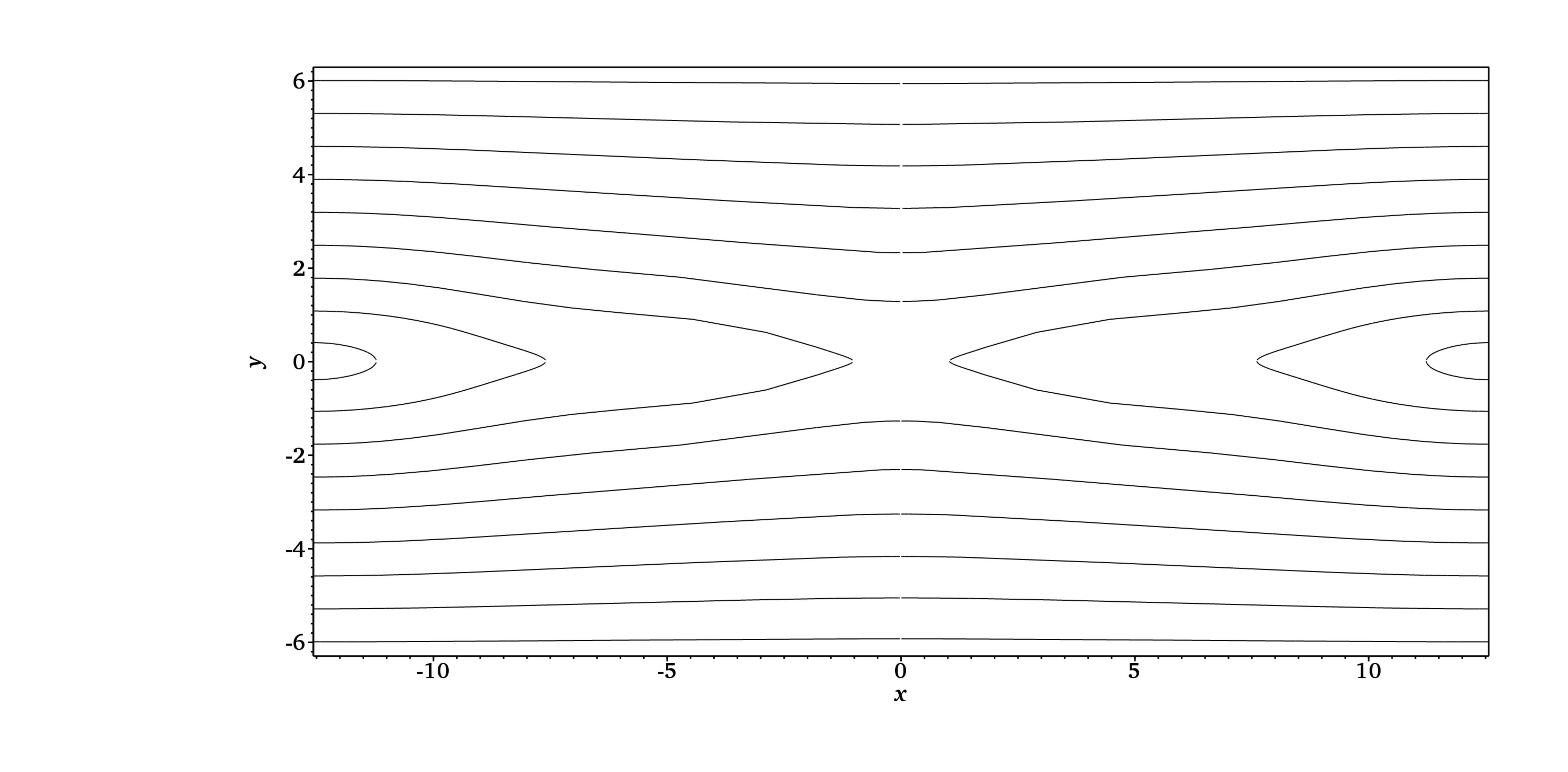}}
  \end{center}
  \caption{Comparison of $j_z$ and magnetic fieldlines between pure Vlasov and coupled simulation.}
  \label{fig:fields}
\end{figure}

\begin{table}
  \begin{center}
    \begin{tabular}{lccc}
      \hline
      Code & System & Resolution & Time \\
      \hline
      multifluid & 32 CPUs & $256\times 128$ & 0.3~h \\
      \hline
      Vlasov (cf. \cite{sch2006}) &  64 CPUs & $256\times 128\times 30 \times 30 \times 30$ & 150~h \\
      \hline
      Vlasov (this work) & 64 CPUs / 64 GPUs & $256\times 128\times 32 \times 32 \times 32$ & 16~h \\
      \hline
      Vlasov/multifluid & 64 CPUs / 64 GPUs & $256\times 128\,(\times\, 32 \times 32 \times 32)$ & 8~h \\
      \hline
    \end{tabular}
  \end{center}
  \caption{Comparison of the computational time required for solving the GEM reconnection challenge problem.}
\label{tab:speed-up}
\end{table}
The results of the coupled simulation look both quantitatively and
qualitatively the same as for the pure Vlasov simulation, which means
that the region in which the determining physical processes occur is
indeed the current layer and the multifluid model is sufficient for
the outer region that was chosen (see Fig.~\ref{fig:gem-curve} and
Fig.~\ref{fig:fields}).

As can be expected, a coupling inside the current sheet does not
yield satisfactory results, because the simple 5-moment model leads to
a different behaviour than the kinetic model. Thus, in future
developments, new strategies will have to be developed in order to identify the regions
which have to be treated by the kinetic or fluid description,
respectively. This would result in an adaptive multiphysics
approach to collisionless plasmas.

One problem, that is observed in general is the heating in the Vlasov
region, due to numerical diffusivity in the velocity space, which can
hardly be avoided. Especially the fast-gyrating electrons tend to be
heated noticeably. This effect does not take place when the
energy conservative multifluid code is used. This gives rise to a
small yet observable temperature gradient at the coupling border and corresponding
changes in other fields affected by it.
As a quick fix, the numerical diffusivity can be lowered by a higher
resolution in $v$-space. In the long run a more robust solution
for this problem needs to be found.

\section{Summary and Outlook}
\label{sec:summary}

A Vlasov code on GPUs and a conventional multifluid code were
presented that can be run both on their own as well as coupled to each other
during runtime. Coupling of both codes was achieved by a combination
of extrapolating and adjusting the distribution function according to
the moments of the fluid description. Specific results on  the GEM
reconnection setup were extremely encouraging and a remarkable
speed-up was observed . However, we should note that this is only a
first step in the treatment of multiphysics descriptions of
collisionless plasmas. Further steps needed to achieve this goal
should include
\begin{enumerate}
  \item[i)] appropriate closure relations (adiabatic, Chew-Goldberger-Low, isotropic, polytropic, Landau-fluid) 
            depending on the physical situation;
  \item[ii)] an identification of regions which have to be treated 
             by a kinetic or fluid description, respectively;
  \item[iii)]  allowing the kinetic regions to change and move in an 
               adaptive way in order to get the most efficient description 
               of the underlying problem (analogous to our our adaptive mesh 
               framework \textit{racoon} \cite{dreher-grauer:2005}).
\end{enumerate}

These topics are currently under development in our group.

\section*{Acknowledgment}

We acknowledge all the fruitful discussions we had with J\"urgen Dreher.
This research was supported by the DFG Research Unit FOR 1048,
project B2. Calculations were performed on the the CUDA-Cluster
DaVinci hosted by the Research Department \textit{Plasmas with Complex
Interactions} at the Ruhr-University Bochum.

\bibliography{lit}

\end{document}